**An immunological autobiography: my year as a COVID-19 vaccine trial participant.**

Ross M. Kedl PhD

Department of Immunology and Microbiology, University of Colorado School of Medicine, Aurora, CO 80045, USA

*Correspondence: ross.kedl@cuanschutz.edu

**Abstract**

I present here longitudinal evaluation of T and B cell immunity to SARS-CoV2 and variants of concern (VOC) from a single subject (me) over an entire year post vaccination. After enrolling in the Moderna phase III clinical trial, I collected my own biological samples pre- and post-immunization in the event of being a recipient of the experimental vaccine. The evidence strongly supports the conclusion that I did not receive the placebo. The analysis is admittedly limited to an n of 1, but the results fit well with data taken from published works and represent one of the more comprehensive longitudinal evaluations of vaccine-elicited immunity within a single individual yet to be undertaken. Though the data amount to a well-documented anecdote, given its granularity, it is not without its insights and may be of further use in directing future longitudinal studies that have actual statistical significance.

## Introduction

The COVID-19 pandemic caused by severe acute respiratory syndrome coronavirus 2 (SARS-CoV-2) is now officially the most devastating pandemic in the US history, at least for the last century. The global response to this threat has been swift, leading to the development of multiple safe and efficacious vaccines in record breaking time. Moderna performed its phase III COVE study of its vaccine, mRNA-1273, at the University of Colorado Anschutz Medical Campus. Being an immunologist who's research focuses on mouse models of vaccine-elicited T cell responses, I enrolled in the trial in order to i) contribute to the process of vaccine approval, ii) potentially gain much-desired immunity against COVID19, and iii) if so, then document my vaccine elicited response in the process. Prior to trial entry, I and my colleagues (see acknowledgements) developed a multiplexed, flow-based method for evaluating SARS-CoV2-specific humoral immunity. This assay included the analysis of both IgG and IgA against the RBD/spike proteins from the original vaccine strain (Wuhan) and the Beta, Gamma and Delta variants of concern (VOC), as well as against 3 seasonal coronavirus strains (HKU1, OC43, and 229E) and tetanus toxoid. In parallel, assays for evaluating antibody-mediated virus neutralization and SARS-CoV2 RBD-specific cellular (CD4 and CD8 T cell) responses were developed as well. With expressed permission from the subject in question (me), I utilized these assays to evaluate multiple biological samples (serum, PBMCs, and nasal swabs) acquired at numerous time points over the course of an entire year following initial vaccination. What follows is (as far as I can tell) one of the more comprehensive longitudinal immunological analyses of a vaccine-elicited response derived from a single individual. The data show time-dependent features of the primary immune response to mRNA-1273 that fit well with published results, and provide some insights into the strength, breadth and durability of immunity after mRNA vaccination.

## Results and discussion.

*Serum evaluation of Innate cytokines reveals elevated IL-1 pre-boost and type II IFN post boost.*

Hearing that the University of Colorado was a site for multiple COVID19 vaccine clinical trials, I applied for enrollment in the first trial to become active on campus, the COVE phase III trial for Moderna's experimental vaccine, mRNA-1273. Upon successful enrollment, and in the event I might receive the vaccine and not the placebo, I began acquiring serum samples immediately before and at numerous time points after my 2 injection regimen. Data from phase I/II Pfizer and Moderna trials indicated a high incidence of short-term side effects (injection site pain, fever, headache, myalgia, etc) post vaccination. I experienced a mild degree of pain approximately 5 hours post injection at the injection site which sustained over the next 3-4 days. As this is not a side effect as commonly associated with a saline injection, this seemed early evidence in favor of the conclusion that I was more than likely not in the placebo group. Evaluation of my serum cytokines found good evidence for this conclusion in the form of greatly elevated IP-10, a highly IFN-sensitive chemokine, at 48 hours post vaccination (**Fig 1A**). Curiously, when evaluated as the fold change in cytokines from pre-vaccine levels, this was the only detectable inflammatory factor (within the panel of cytokines evaluated) after my initial vaccination (**Fig 1B**), perhaps explaining my lack of any additional symptomology. I also took serum samples just before and after my boosting injection 28 days later. When normalized to the cytokine levels found in the pre-primary vaccination serum sample, three features of my innate signature surrounding the second injection were of interest. First, IL-1beta and IL-1ra were elevated at 28 days before the second injection (**Fig 1C**). These results suggest the potential of ongoing inflammasome activation (and concomitant IL-1 production) after the priming dose, forming the biological basis for the fever that is more often experienced by vaccinees (though curiously, not me) after the secondary vaccination. Second, even more IP-10 was observed at 48 hours post boost, potentially indicating even greater amounts of type I IFN produced after the boost than the priming injection (**Fig 1C**). As IFN is an innate cytokine for which any kind of "memory" is not usually anticipated, this increase in IFN was the result of either some version of "trained immunity" or, more likely, the increased presence of inflammatory cells within the injection site (which for me was the same for both injections). Third, this elevated IP-10 could have also been influenced by an unexpected and substantial spike in IFN$\gamma$ seen at 24 hours post boost (**Fig 1C**). Given the fact that this was unique to the secondary vaccination, it may be the result of NK cell activation mediated by Fc receptor crosslinking by anti-RBD antibody formed after the first vaccination (see below). However, a role for CD4/CD8+ T cells in this elevated IFN$\gamma$ signature cannot formally be ruled out (see Figure 4). Regardless, the detection of IFN$\gamma$ tracked well with the moderate-to-significant myalgia and headache I experienced after the second injection (objectively, the second shot packs a wallop). Collectively, these results agree well with those published by Pulendran and colleagues (Arunachalam et al., 2021) and support a biological basis for why the majority of individuals receiving the mRNA vaccines experienced fever (IL-1) and/or myalgia/headache (IFN$\gamma$) specifically associated with the second vaccination.

*Vaccine-elicited SARS-CoV2-specific IgG and IgA correlate with virus neutralization and predict the observed ~8-month window for the waning of vaccine-elicited immunity.*

Early in the pandemic, I and my colleagues developed a multiplexed assay for the purposes of evaluating SARS-CoV2 specific humoral immunity (Sabourin et al., 2021; Schultz et al., 2021). The assay was eventually expanded to include the quantification of IgG and IgA against SARS-CoV2-RBD, related SARS-CoV2 VOCs spike proteins, and three seasonal strains of coronavirus. Proteins of interest conjugated to BioLegend LegendPlex 5 and 7um carboxylated microspheres bearing different levels of APC fluorescence (**Fig 2A**) serve as the substrate for flow-based detection of IgG and IgA specific for each protein (**Fig 2B**). Because the fluorescence intensity of the detecting fluorophore is proportional to the amount of IgG or IgA bound to each bead, the magnitude of the response at each time point can be comparatively evaluated and stratified using each bead's geometric mean fluorescence intensity (gMFI) (**Fig 2C**). Overlaying histograms showed that I acquired detectable anti-SARS-CoV2 RBD IgG between 7 and 10 days post immunization. The amount of antibody from the primary immunization peaked at 21 days, and the gMFI increased ~8 fold by 11 days after the secondary injection (**Fig 2C**). Thus, the time to the peak of immunity after secondary antigen challenge is approximately half of that needed to achieve the peak after the initial vaccination.

I performed serial dilutions of serum from each time point for IgG (**Fig 2D**) and IgA (**Fig 2E**) and quantified the response using the 50% midpoint of the effective antibody concentration (EC50, **Fig 2F**). The very fact that IgA responses were so robustly induced came initially as a surprise, as intramuscular vaccines are not known for their capacity to induce this isotype. That said, others have noted its production after mRNA-based vaccination (Corbett et al., 2021; Isho et al., 2020; MacMullan et al., 2021; Nahass et al., 2021). In my case, IgA levels were ~20 fold less than IgG (**Fig 2F**). Interestingly, between 150-200 days, the IgG EC50 declined to the same level observed after the initial vaccination (**Fig 2F,** dotted lines). This time frame (~6 months) seems consistent with the waning of optimal protection against breakthrough infections. As I am well under 65, in good health and without any COVID risk factors, these results seem to make booster vaccines for anyone 6 months out from their original vaccination an advisable approach.

My colleagues evaluated virus neutralization in each serum sample, identifying the dilution of antibody necessary to achieve a 50% reduction of focus formation in a focus reduction neutralization test (FRNT50). The degree of virus neutralization tracked exceptionally well with both the gMFI of anti-RBD IgG (**Fig 2G**) as well as the calculated EC50 (**Fig 2H**) at each time point. Indeed, the correlation between each of these immunological parameters was highly predictive of neutralization capacity (**Fig 2 I and J**), consistent with published data on this correlation (Corbett et al., 2021; Krammer et al., 2021; Mateus et al., 2021; Wajnberg et al., 2020). Interestingly, my virus neutralization capacity appeared slightly more stable than its EC50, dropping below my initial priming level nearly 100 days after the same decline in total IgG (**Fig 2J**). Further, the fold change in virus neutralization after the booster immunization substantially outpaced that seen for total IgG and IgA. Thus, while total IgG increased 3-5-fold after the second dose, virus neutralization capacity was augmented nearly 30-fold (**Fig 2L**). Collectively, these data indicate that the second injection favors an increased overall quality of the antibody responses, consistent with the process of affinity maturation as previously noted (Pratesi et al., 2021; Sette and Crotty, 2021; Turner et al., 2021).

*Vaccine-elicited immunity against VOCs, oral/nasal immunity, and background responses to seasonal strains of coronavirus.*

As the pandemic progressed, different VOCs became the dominant circulating strains, and as such became the focus of analysis for the degree of cross reactivity for vaccine-elicited immunity. Our multiplexed assay facilitated incorporation of these VOCs into the longitudinal analysis of the antibody response (**Fig 2A,B**). Determining my VOC-specific immunity revealed reduced reactivity to the VOCs as compared to Wuhan RBD, with a similar overall pattern of immune progression; VOC-specific antibody responses showed a 4-6 fold increase after the second dose followed by a similar waning trajectory (**Fig 3A and B**). As with immunity to Wuhan RBD, my Delta-specific immunity dropping below that observed after the initial vaccination somewhere after ~150 days, (**Fig 3A**). Given the correlations between virus neutralization and EC50 for the Wuhan strain (**Fig 2K**), these data are again consistent with the increased susceptibility to a breakthrough infection observed for Delta ~6 months post vaccination (Christensen et al., 2021; Eyre et al., 2021), further supporting the need for booster vaccines based on time post vaccination and not qualified by age or other risk factors.

I also evaluated my antibody responses to 3 seasonal strains of coronavirus and to tetanus toxoid. These responses provide insights into the relationship between SARS-CoV2-specific immunity relative to other infections or vaccinations. As it turns out, I had detectable IgG (**Fig 2B and 3B**) and IgA (not shown) to all three seasonal strains of coronavirus prior to my mRNA-1273 vaccination. mRNA-1273 vaccination did not demonstrably augment my overall antibody responses to the 229E and HKU1 seasonal strains (**Fig 3B and C**), consistent with there being no cross reactivity between these strains and the SARS-CoV2 spike protein. This point was further emphasized when I acquired flu-like symptoms after attending a scientific conference ~300 days after my initial vaccination. I reported my symptoms to the COVE clinical trial coordinators, was evaluated for COVID19 and found to be negative. The trial provided additional PCR-based evaluation for ~20 other viral infections, and (in a fit of irony) I was found to have contracted the non-pandemic 229E seasonal strain of coronavirus in the middle of a coronavirus pandemic. My serum antibody levels specific for 229E reflected this by increasing substantially (**Fig 3B and C**). However, this did not result in any change in antibody titers (IgG or IgA) against SARS-CoV2 (**Fig 3A**). I also received a Tdap booster ~100 days post mRNA-1273 vaccination. As with 229E infection, my antibody titers against TT demonstrably elevated (**Fig 3B**) but my SARS-CoV2 specific immunity was again unaffected **Fig 3A**).

Curiously, the impact of mRNA-1273 vaccination on IgG specific for the OC43 seasonal coronavirus strain was quite different. OC43-specific IgG was substantially elevated immediately after my vaccination, most easily seen as an increase in the gMFI of OC43-specific IgG (**Fig 3C**), though also observable in the EC50 (**Fig 3B).** This suggested some degree of cross-reactivity between existing OC43 immunity and the SARS-CoV2 spike protein encoded in the mRNA-1273 vaccine. Previous reports indicated increased cross reactivity between the SARS-CoV2 S2 domain and seasonal coronaviruses (Kaplonek et al., 2021; Ng et al., 2020). More remarkably, Alter and colleagues observed that pre-existing responses to OC43 predicted earlier development of SARS-CoV2 immunity and decreased severity of COVID19 infection (Kaplonek et al., 2021). We therefore modified the multiplex assay to examine antibody reactivity to the S2 domain of the SARS-CoV2 spike protein and evaluated the initial time points (0-92 days) over which elevated OC43 antibody responses were found. Indeed, a robust correlation was found between my SARS-CoV2 S2-specific IgG and the OC43 spike IgG (**Fig 3D**). In contrast, no such correlation was found between HKU1 and SARS-CoV2 S2. These data support the conclusion that the SARS-CoV2 S2 domain shares sufficient similarity with the S2 domain from OC43 such that mRNA-1273 vaccination augments preexisting immunity to OC43.

I also examined saliva and nasal swab samples for the presence of anti-SARS-CoV2-specific IgG and IgA, as published data indicated that antibodies could indeed be found in these sites following mRNA vaccination (Corbett et al., 2021; Isho et al., 2020; MacMullan et al., 2021; Nahass et al., 2021). Given my seasonal coronavirus infection, intranasal IgG and IgA against 229E seemed an effective positive control for the identification of mucosal antibodies. My results revealed a high amount of intranasal SARS-CoV2-specific IgG and IgA (**Fig. 3E**). While intranasal 229E-specific IgA was elevated compared to SARS-CoV2-specific IgA, the amount of vaccine-elicited SARS-CoV2-specific IgG was considerably superior to 229E-specific IgG. It is unclear as to whether the anti-SARS-CoV2-specific IgG and IgA found in the mucosa was derived from intra- or extra-mucosal antibody production, though spill-over from extra-mucosal sources seems most likely. Regardless, earlier notions that the mRNA-vaccines failed to generate immune protection within the respiratory mucosa clearly need re-evaluation, particularly in light of the results from Seder and colleagues who identified respiratory-based immunity as a primary correlate of vaccine-elicited host protection in non-human primates (Corbett et al., 2021).

*Tertiary antibody responses to third vaccine dose.*

One year in, Moderna incorporated a 3rd immunization into their trial design, and I was given the tertiary booster injection 407 days after my initial vaccination. Within 7 days post-boost, antibodies (IgG) specific to both the original Wuhan strain as well as to the Delta Variant improved substantially, extending beyond even the peak response observed after the secondary vaccination (**Fig 4A**). The tertiary response peaked between 12-16 days post boost, and then began a decline more gradual than that observed after the second dose, as evaluated by the gMFI at a single serum dilution (**Fig 4A, left)**, and after calculation of EC50 from serial dilutions at each time point **(Fig 4A, right).** This difference in antibody decay between post-secondary and -tertiary immunizations was most easily observed by evaluating the decay of the secondary and tertiary antibody responses by curve fit. The decline of anti-RBD IgG after the secondary vaccination fit exceptionally well ($R^2$=0.9957) to an exponential, one phase decay rate (**Fig 4B**). In contrast, anti-RBD IgG declined in a strictly linear fashion ($R^2$=0.9994) after the tertiary immunization (**Fig 4B**). Thus, not only is the peak antibody response higher post tertiary vaccination, the waning of antibody over time operates as an arithmetic, not geometric regression. While the loss of antibody after the secondary vaccination took ~150 days to fall to that observed after the initial vaccine dose (day 21-28), the arithmetic decay of the tertiary response predicts taking ~250 days to reach the same post-primary vaccine peak. In addition to this increase in the durability of the vaccine response against the original Wuhan-derived RBD, tertiary vaccination had an even more substantial impact on the breadth of antibodies reactive to the Delta and Beta variant spike proteins **(Fig 4A)**. This was best observed by comparing the fold-change in antibody between the peaks of the primary and secondary vaccinations (**Fig 4C, "secondary"**) to the fold change in antibody pre- and post-tertiary vaccination **(Fig 4C, "tertiary")**. Though the overall magnitude of antibody specific for the Wuhan strain was the greatest (**Fig 4A**), the fold change was considerably higher for two VOCs (**Fig 4C**). Thus, despite utilizing the same RBD sequence derived from the original Wuhan strain, tertiary immunization elicited an antibody response of greater magnitude **(Fig 4A)**, durability **(Fig 4B)** and breadth **(Fig 4C)** than the secondary vaccination. These results are highly consistent with the observation that boost vaccination generates increased protective immunity even against the most recent variants of concern such as Delta and Omicron (Garcia-Beltran et al., 2021; Nemet et al., 2021).

*T cell effector function peaks shortly after vaccination and drops to a stable memory pool.*

My PBMCs obtained at various times post vaccination were stimulated by overlapping peptides covering the RBD domain of SARS-CoV2 (Davenport et al., 2021) and evaluated using the Activation Induced Marker (AIM) assay as described by Sette, Crotty and colleagues (Sette and Crotty, 2021; Tarke et al., 2021). Antigen-responsive T cells were identified as dual CD69+CD137+ 24 hours post peptide stimulation (**Fig 5A, C**). Within this subset of cells (AIM+) I also evaluated the frequency of my cells producing cytokines (**Fig 5B, D**). While the total number of CD4+ T cells peaked later than the total number of Aim+ CD8 T cells (**Fig 5C**), the frequency of IFNg+ in both CD4 and CD8 T cells was highest ~4 weeks after the second vaccination. The remaining longitudinal samples indicated a declining frequency of effector (AIM+IFNg+) (**Fig 5D**), but relatively stable population of total **(Fig 5C**), SARS-CoV2-specific T cells over the course of the year. My results are again consistent with published data showing that the second vaccination generates an elevation in effector T cells (Goel et al., 2021; Oberhardt et al., 2021) but does not seem to compromise the generation and maintenance of long lived memory T cell frequencies 6-8 months post immunization (Goel et al., 2021; Mateus et al., 2021; Oberhardt et al., 2021). My results also add emphasis to the importance of using T cell detection assays independent of specific effector functions (other than surface marker expression); using IFNγ production as the sole identifier of antigen specificity would not only have underestimated the frequency of my SARS-CoV2-specific T cells post vaccination, it would have also indicated a gradual decline in that frequency over time. At the time of the writing of this manuscript, post-boost ($3^{rd}$ dose) T cell analysis had yet to be performed.

Care must of course be taken in applying my results to that of the broader public. That said, longitudinal evaluation of a single response can provide insights for its broader applicability, particularly when the data found connects well with published observations as does mine here. For example, a general correlation between the amount of IgG and virus neutralization has been noted (Sadarangani et al., 2021; Sette and Crotty, 2021) and my results suggest that this correlation within a specific individual may well be even better than previously appreciated.

Similarly, Galit and colleagues showed that immunity to OC43 correlated with reduced severity of COVID19 (Kaplonek et al., 2021). One might anticipate this conclusion to be limited by being derived from COVID19 patients for whom sample acquisition could only occur post infection. My results clearly indicate a rise in antibodies i) uniquely reactive to OC43 among the seasonal strains, ii) dependent on mRNA-1273 vaccination, and iii) correlating exceptionally well with antibodies against the S2 domain. The fact that this was seen after being vaccinated only against the spike protein adds further strength to the conclusion that OC43 cross reactivity to the SARS-CoV2 S2 domain is the source of the biological phenomenon observed.

The data presented here provide an effective timeline for the durability of what one might consider an "average" vaccine response for the first round of vaccinations. My data reinforce the 5-7 month time point for the waning of humoral immunity (~5 for total IgG, ~7 for neutralization titers) after the primary rounds of vaccination. These data suggest that booster doses would be more appropriately timed in the direction of the 5-month time point, a fact that the CDC affirms as well (https://www.cdc.gov/media/releases/2022/s0104-Pfizer-Booster.html). One of the most compelling and encouraging features of the data are the differences between secondary vs tertiary responses. 3rd dose immunization enhances the magnitude and breadth of the antibody response that available data from other groups indicates is sufficient for mediating reasonable levels of virus neutralization even against strains as diverse as Omicron (Garcia-Beltran et al., 2021; Nemet et al., 2021). Additionally, the durability of the tertiary response is greatly augmented relative to that of the secondary, with a linear rather than exponential decay rate.

The history of science is full of examples in which good use has been made from the experience of, and/or data derived from, that of one or two subjects. The history of vaccinology is no different, an excellent example being the initial establishment of suitable anti-tetanus toxin titers following toxin challenge of just two vaccinated individuals (https://www.nvic.org/vaccines). The data presented here was acquired, and is presented, in that spirit. Beyond being an interesting way for an immunologist to keep occupied during a pandemic, the longitudinal granularity of these data may have some utility to future evaluations of vaccine-elicited responses.

**Acknowledgements**

A data set such as this truly takes a village to acquire. I am grateful to the following "village" of collaborators: Rosemary Rochford PhD for finding better ways of using the MMI than just analyzing my own serum samples; Tem Morrison PhD, Mary McCarthy PhD and Bennett Davenport PhD for establishing the FRNT50 assay for the benefit of so many on campus over the course of the pandemic; Ryan Baxter, Berenice Cabrera-Martinez, and Elena Hsieh MD for their dedication to developing and implementing the T cell assay for RBD-specific CD4 and CD8 T cells; Ashley Frazer-Abel PhD for analysis of innate cytokines; Jared Klarquist PhD for generating our own peptide library for the analysis of mouse epitopes for SARS-CoV2 RBD; Cody Rester for managing the MMI, and pretty much everything else.

**Materials and Methods**

*Immunizations and sample collection*

The subject enrolled in the Moderna COVE phase III trial and received his first injection on August 25, 2020. In the event he received the vaccine and not the placebo, the subject began collecting blood samples starting the morning before vaccination and at numerous time points afterward as indicated in the text. Serum was isolated using BD SST Vacutainer tubes. For T cell assays, peripheral blood samples were collected in sodium heparin and processed promptly to isolate peripheral blood mononuclear cells (PBMCs) which were frozen in liquid nitrogen until use. Nasal swabs and saliva were also taken at various times post vaccination.

*Milliplex Luminex measurements of Ten-Plex of Inflammatory Cytokines.* Innate cytokines were evaluated by Exsera BioLabs utilizing the Luminex MagPix technology and the Millipore's MILLIPLEX® MAP reagents kits according to the manufacturer's instruction and run in compliance with SOP.EXS.028 Milliplex Assays for Complement and Cytokine Proteins, SOP.EXS.008 Assay Acceptance and Rejection Criteria and all other applicable SOPs.

*Multiplexed Microsphere Immunoassay (MMI).* An MMI was developed using Biolegend caboxylated LegendPlex microbeads to simultaneously quantify IgG and IgA against the spike RBD and nucleocapsid of the Wuhan strain of SARS-CoV-2, three VOCs (beta, gamma, delta), three season coronavirus strains (OC43, 229E, HKU1), and tetanus toxoid (TT) as a positive control. Bovine serum albumin (BSA) conjugated beads were used as a negative control. All SARS-CoV2 and seasonal coronavirus proteins were obtained from either BEI or Sino Biological.

TT was obtained from Millipore. Multiplex bead protein conjugation, sample incubation and flow cytometric analysis will be performed as previously described (Sabourin et al., 2021; Schultz et al., 2021). Geometric mean fluorescence intensity (gMFI) of the IgG/IgA for each sample and dilution was captured with a CytoFLEX S flow cytometer (Beckman Coulter) and analyzed with FlowJo (version 10.7.1; BD Biosciences). Prism (version 8.4.3, GraphPad) was used to plot data.

*SARS-CoV-2 Ab-mediated neutralization assay (focus reduction neutralization test, FRNT):* FRNT assay was performed as previously described (Schultz et al., 2021). Briefly, samples were heat-inactivated and serially diluted (starting at 1:10) into microwells (96 well plate). Approximately 100 focus-forming units of SARS-CoV-2 USA-WA1/2020 was added to each well and the serum plus virus mixture incubated for 1 h at 37°C prior to addition to cells. After 2 h, samples were removed, cells overlaid with 0.5% methylcellulose and incubated 30 h at 37°C. Cells were fixed with 4% paraformaldehyde and probed with 500 ng/mL of an anti-SARS-CoV spike monoclonal Ab (CR3022). Foci were detected using horseradish peroxidase-conjugated goat anti-human IgG, visualized with TrueBlue substrate and counted using a CTL Biospot analyzer and Biospot software.

*SARS-CoV-2 specific T cell assessment via Activation Induced Markers (AIM) assays:* We constructed our own peptide library (Davenport et al., 2021), comprised of 15-mer peptides and overlapping by 11 amino acids, covering the entire SARS-CoV-2 Wuhan-Hu-1 RBD sequence (RBD; GenBank identifier: MT380724.1). This library was used to perform the AIM T cell assay as previously described (Mateus et al., 2021; Tarke et al., 2021). Briefly, PBMCs were thawed, rested overnight, and stimulated for 24 hours with 2 ug/ml of the RBD peptide pool and 1 ug/mL of anti-human CD28/CD49d (BD Biosciences). Unstimulated samples were treated with co-stimulation alone, PHA was used as a positive control for PBMC viability and functional response to T cell stimulation. Brefeldin A was added after 20 h post stimulation for 5 hours at 37°C to capture intracellular cytokine production. After 5 hours, cells were stained with antibodies to CD4, CD8, CD69, CD137, CCR7 (to exclude naïve), CD19 (to exclude B cells) and ghost dye (for live/dead exclusion). Cells were permeabilized and stained with antibodies to IFNγ and TNFα. After background-subtraction using paired unstimulated control samples, AIM+ cells were identified by dual expression of CD69 and CD137. From AIM+ CD4 and CD8 T cells, intracellular IFNγ and TNFα production was subsequently evaluated. Flow cytometry data were acquired on a four-laser (405, 488, 561, 638 nm) CytoFLEX S flow cytometer (Beckman Coulter) and analysis was performed using FlowJo (version 10.7.1; BD Biosciences). Prism (version 8.4.3, GraphPad) was used to plot data.

## Figure Legends

**Figure 1. Innate cytokines in response to primary and secondary mRNA-1273 vaccination**

Innate cytokine levels were evaluated by Luminex in the serum samples obtained at the time points indicated. A) total cytokine amounts before initial vaccination (pre) and at 24 and 48 hours post first injection. B) all cytokine levels normalized to pre-vaccination levels and expressed as fold change. C) Similarly normalized data covering both primary and secondary injections. Only the cytokines with a demonstrable fold change from baseline (pre-primary vaccination) are shown.

**Figure 2. One year of IgG and IgA antibody titers and antibody-mediated virus neutralization post mRNA-1273 vaccination**.

A) representative dot plot of the MMI showing both 5 and 7 um beads, their level of APC fluorescence, and the proteins conjugated to each. B) example data showing both PE (anti-IgG) and FITC (anti-IgA) staining of each bead after incubation with serum derived from pre- and day 42 post-vaccine samples followed by anti-hIgG-biotin/SA-PE and anti-hIgA-FITC. C) Histogram offset overlay of anti-RBD IgG staining from day 0-42 post vaccination serum samples diluted at 1:2000. D and E) example serial dilutions for IgG (D) and IgA (E) for samples obtained at the indicated time points. F) The curves in D and E were used for determining EC50. G and H) FRNT50 values for each time point plotted in parallel to anti-RBD IgG gMFI (G) or EC50 (H). I and J). Both the gMFI and EC50 were plotted against FRNT50 values for each time point. The $R^2$ of the correlation and the statistical significance for each is shown. Numbers indicate the time point of each data point K) Maximal fold change in IgG and IgA EC50 and FRNT50, between day 21 and the peak of each response post-secondary vaccination (day 42 for IgG and IgA, day 56 for FRNT50).

**Figure 3. Antibody durability to Variants of Concern and seasonal strains of coronavirus post mRNA-1273 vaccination**.

A and B) IgG titers as measured by EC50 against Wuhan-RBD and spike proteins from the Beta (S. Africa), Gamma (Brazil) and Delta (India) VOCs (A), or against TT and the seasonal coronavirus strains 229E, HKU1 and OC43 (B). C) gMFI of IgG against seasonal coronavirus strains at a serum dilution of 1:200. D) Correlation of S2-specific IgG gMFI with IgG gMFI for OC43 but not HKU1. Graph shows the gMFI for each time point from day 0-92 at both 1:100 and 1:500 serum dilutions. E) Nasal swabs were eluted in 500ul of buffer, saliva was filtered, and both evaluated for IgG and IgA at a 1:2 dilution. Serum comparison was at 1:2000 dilution.

**Figure 4. Comparison between secondary and tertiary antibody response after 3rd dose boost**.

A) Wuhan, Delta and Beta RBD-specific IgG levels as measured by gMFI (left) or EC50 (right) after all vaccine doses indicated by arrow. gMFI is shown from 1:2500 dilution of serum samples. (B) Decay of antibody levels (from EC50 in A) in the days after the second (blue circles) or third vaccine dose (green triagles). Blue lines track one phase decay (for secondary) or linear (for tertiary) EC50 curve fit.

**Figure 5. Longitudinal analysis of T cell responses following mRNA-1273 vaccination**.

PBMCs were thawed, rested overnight, then stimulated for 24 hours with a peptide library of 15 mer peptides covering the entire RBD domain with a 11 amino acid overlaps. For the last 5 hours, brefeldin A was added. Cells were then washed, surface stained, fixed, permeabilized, and stained for intracellular cytokines. A) T cells at 56 days post vaccination analyzed by flow cytometry for CD69 x CD137 double positive cells within all live, CD19-, non-naïve (CCR7-), CD4 (top row) or CD8 (bottom row) events. Numbers above gate reflect the percentage of total non-naïve CD4 or CD8 events. B) Intracellular cytokine staining of the cells shown on A. C) quantification of events as shown in A) over time and calculated as # of $CD69^+$ x $CD137^+$ cells per 1 million PBMCs. D) Quantification of $AIM^+IFN\gamma^+$ events as shown in B) over time.

# Figure 1

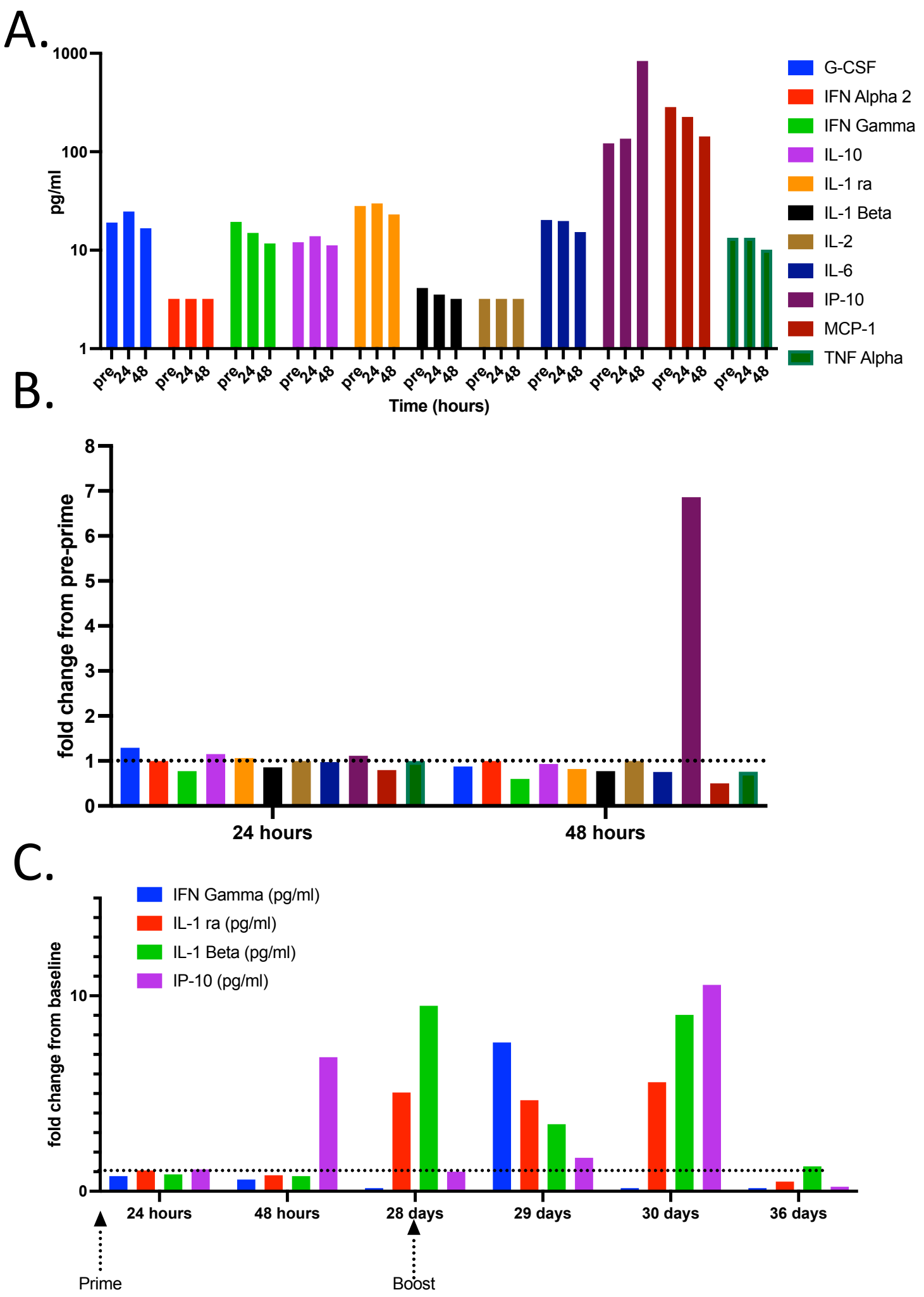

Figure 2

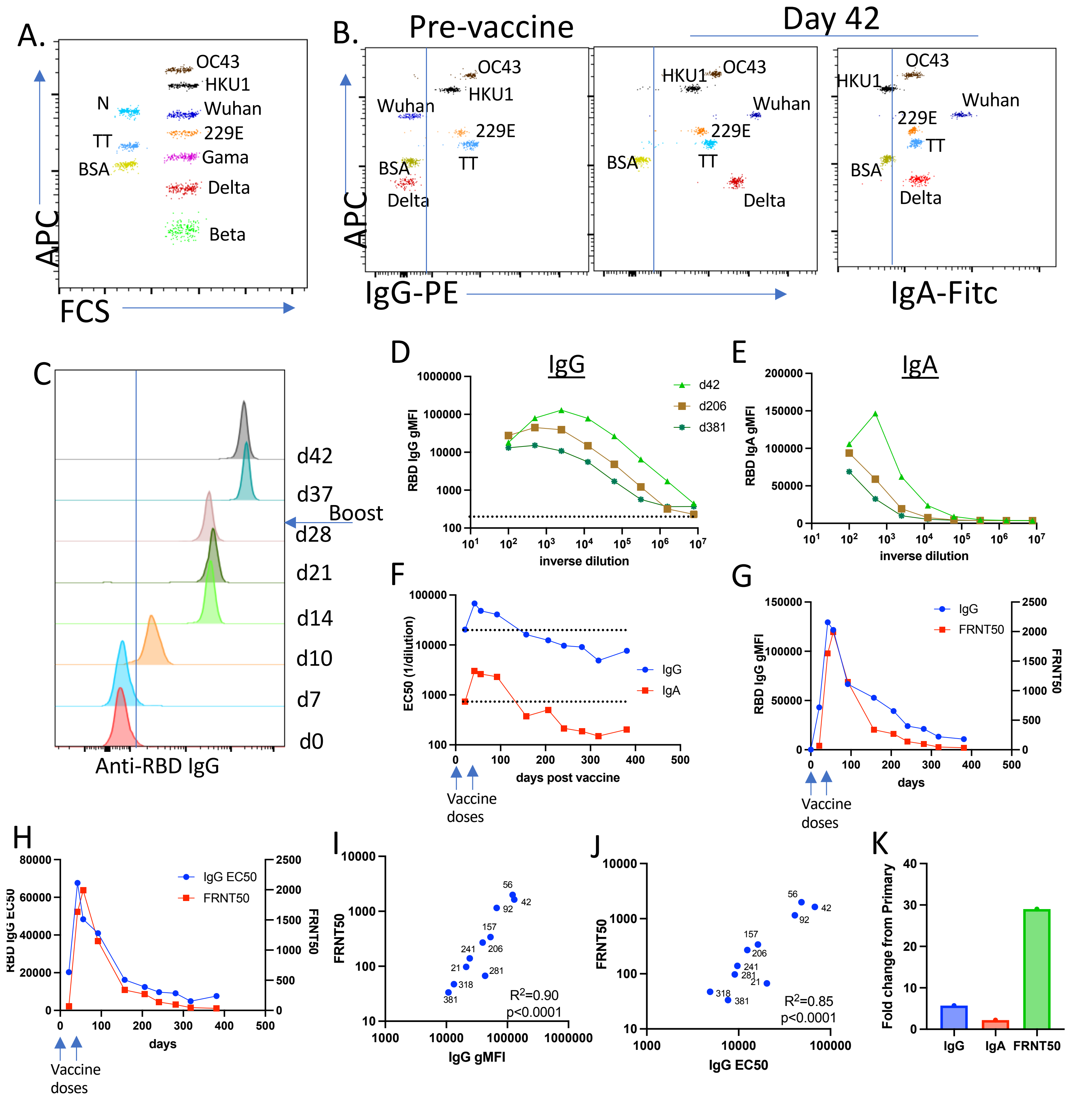

# Figure 3

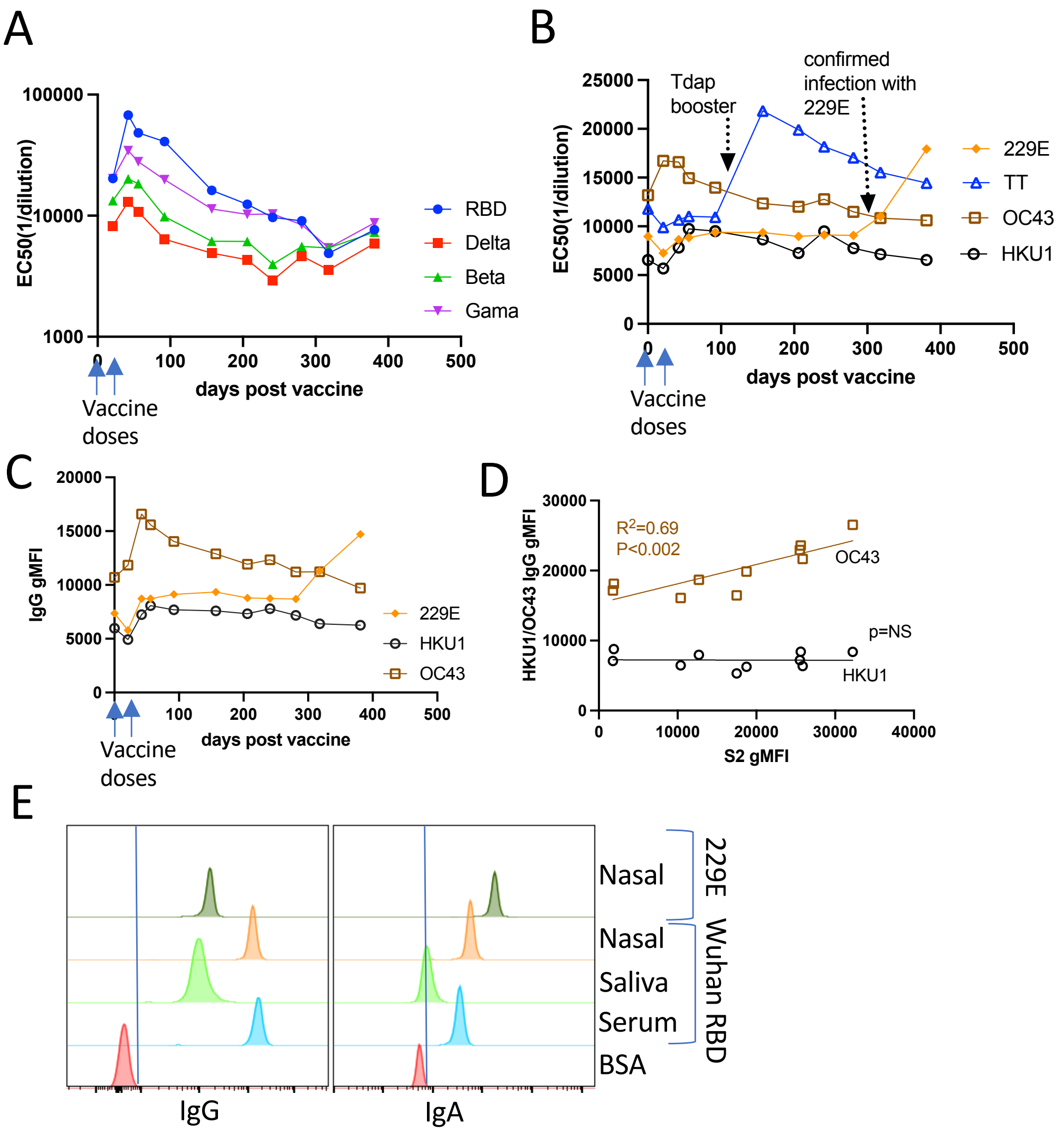

Figure 4

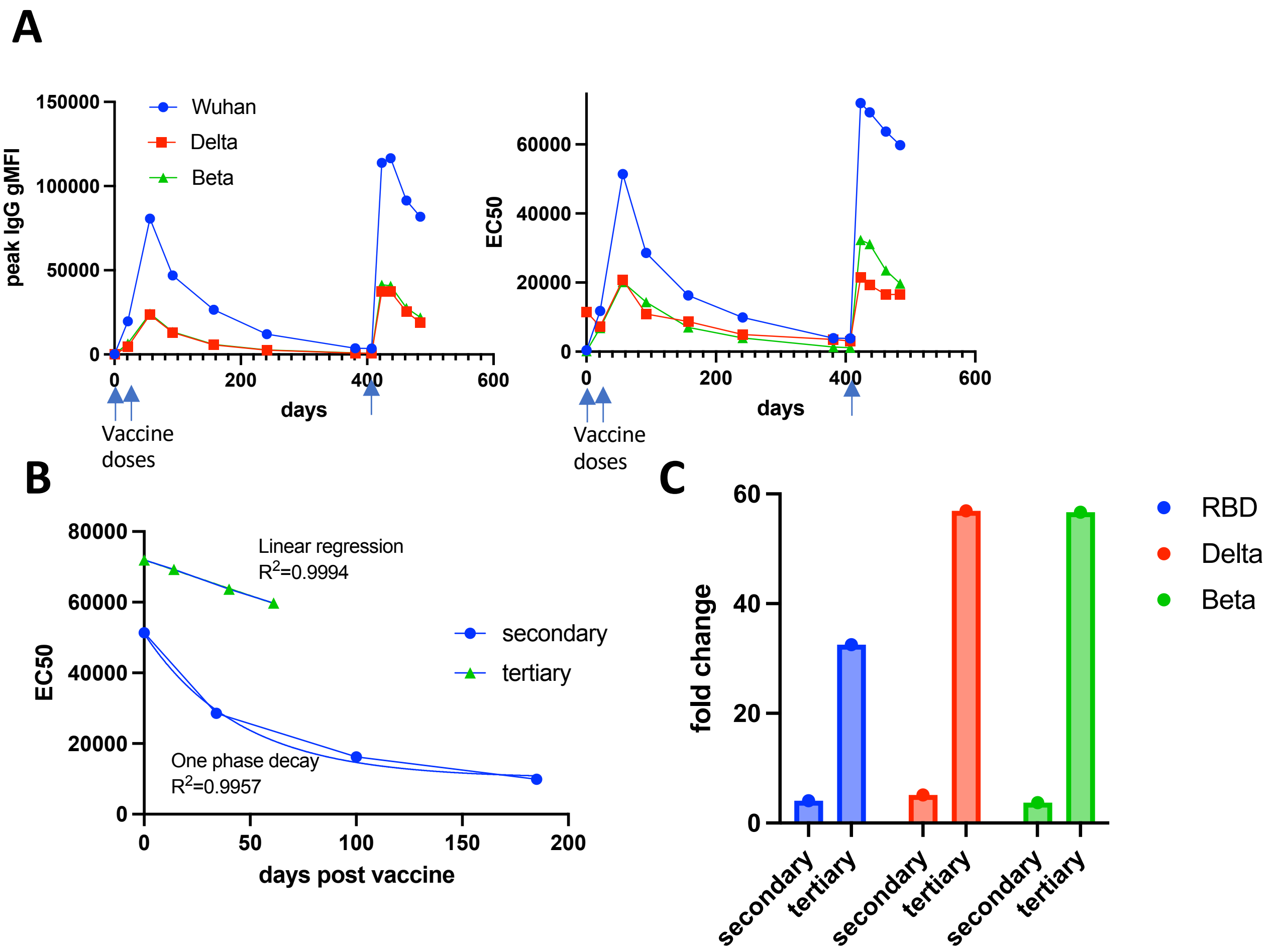

Figure 5

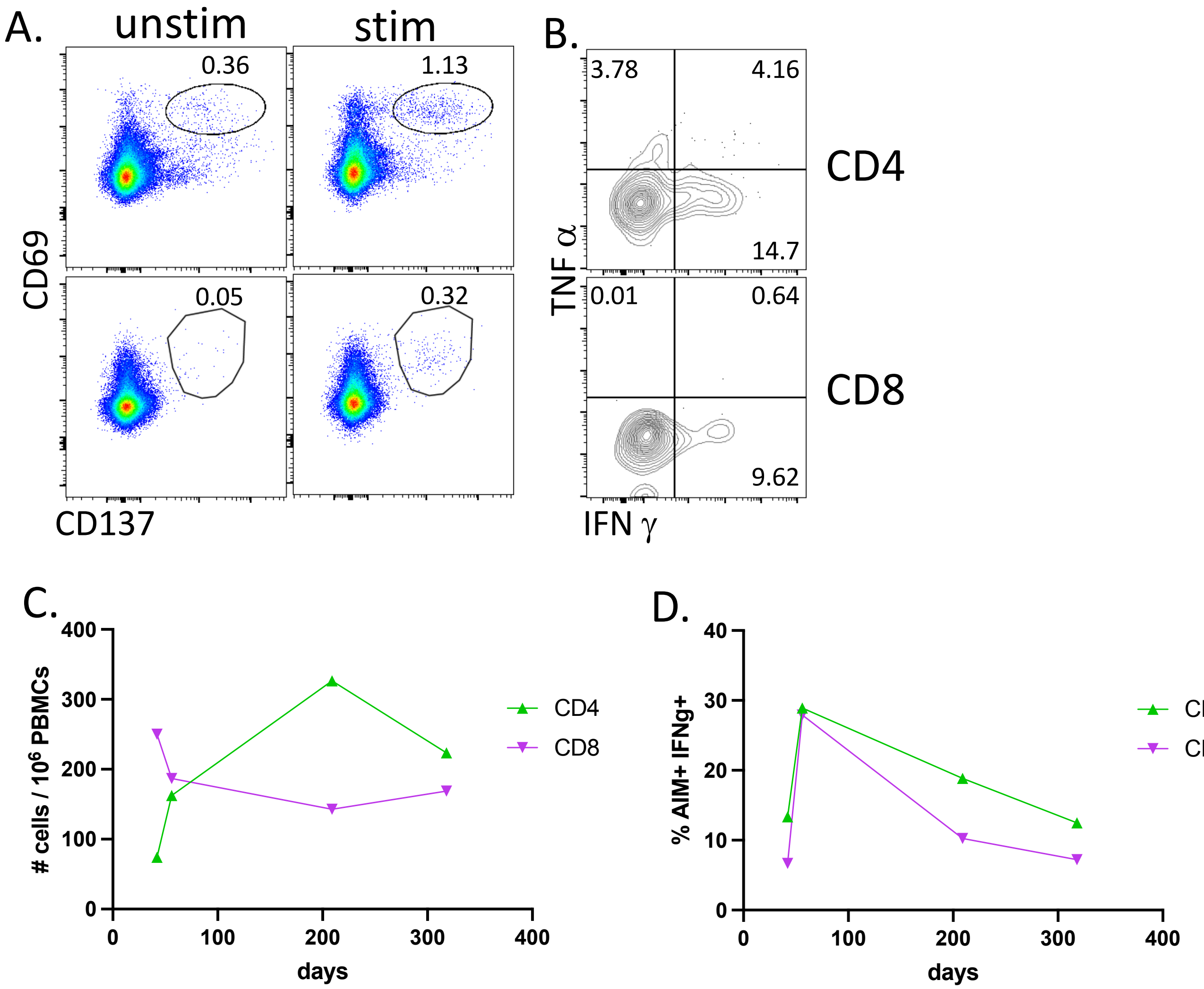